\documentclass{article}
\usepackage{spadre2008}
\usepackage{graphicx}
\frompage{000} \topage{000}                                              
\def\rightmark{Constraining the symmetry energy and its astrophysical impact}
\def\leftmark{Bao-An Li et al.}

\title{Constraining the density dependence of nuclear symmetry energy with heavy-ion reactions
and its astrophysical impact}
\authors{
{Bao-An Li$^1$, Lie-Wen Chen$^{2}$, Che Ming Ko$^{3}$, Plamen G. Krastev$^{1}$ and Aaron Worley$^{1}$%
}\\[2.812mm]
{\normalsize
\hspace*{-8pt}$^1$ Department of Physics, Texas A\&M University-Commerce,\\
Commerce, Texas 75429-3011, USA\\[0.2ex]
\hspace*{-8pt}$^2$ Institute of Theoretical Physics, Shanghai Jiao
Tong University,\\ Shanghai 200240, P.R. China\\[0.2ex]
\hspace*{-8pt}$^3$ Cyclotron Institute and Department of Physics,
Texas A\&M University,\\
College Station, Texas 77843-3366, USA\\[0.2ex]
}}
\abstract{Recent analyses of several isospin effects in heavy-ion
reactions have allowed us to constrain the density dependence of
nuclear symmetry energy at sub-saturation densities within a
narrow range. Combined with constraints on the Equation of State
(EOS) of symmetric nuclear matter obtained previously from
analyzing the elliptic flow in relativistic heavy-ion collisions,
the EOS of neutron-rich nuclear matter is thus partially
constrained. Here we report effects of the partially constrained
EOS of neutron-rich nuclear matter on the mass-radius correlation,
moment of inertia, elliptical deformation and gravitational
radiation of (rapidly) rotating neutron stars.}


\begin{document}

\maketitle
\setcounter{page}{1}

\section{EOS of neutron-rich nuclear matter partially constrained by
heavy-ion reactions}\label{intro}

\begin{figure}[htp]
\centering
\includegraphics[scale=0.8]{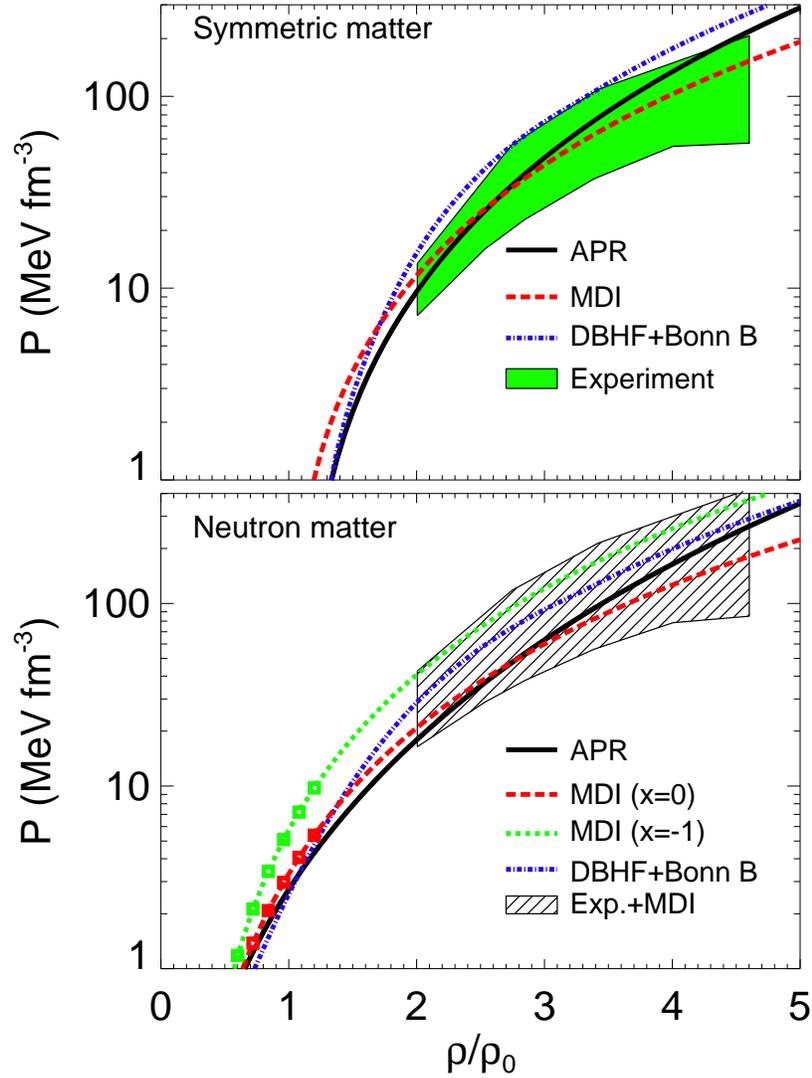}
\caption{(Color online) Pressure as a function of density for
symmetric (upper panel) and pure neutron (lower panel) matter. The
green area in the upper panel is the experimental constraint on
symmetric matter. The corresponding constraint on the pressure of
pure neutron matter obtained by combining the flow data and an
extrapolation of the symmetry energy functionals constrained below
$1.2\rho_0$ by the isospin diffusion data is the shaded black area
in the lower panel. Results taken from Refs.\
\protect\cite{dan02,Kra08b}.} \label{Pressure-eos}
\end{figure}

The EOS of isospin asymmetric nuclear matter can be written within
the well-known parabolic approximation as
\begin{equation}  \label{ieos}
E(\rho ,\delta )=E(\rho ,\delta =0)+E_{\rm sym}(\rho )\delta
^{2}+\mathcal{O} (\delta^4),
\end{equation}
where $\delta\equiv(\rho_{n}-\rho _{p})/(\rho _{p}+\rho _{n})$ is
the isospin asymmetry with $\rho_n$ and $\rho_p$ denoting,
respectively, the neutron and proton densities, $E(\rho ,\delta
=0)$ is the EOS of symmetric nuclear matter, and $E_{\rm
sym}(\rho)$ is the density-dependent nuclear symmetry energy. The
latter is very important for many interesting astrophysical
problems~\cite{lat01,Ste05}, the structure of rare
isotopes~\cite{brown} and heavy-ion
reactions~\cite{ireview98,ibook01,dan02,ditoro,LCK08}. However,
the density dependence of the nuclear symmetry energy has been the
most uncertain part of the EOS of neutron-rich matter.
Fortunately, comprehensive analyses of several isospin effects
including the isospin diffusion~\cite{Tsa04,Liu07} and
isoscaling~\cite{She07} in heavy-ion reactions and the size of
neutron skin in heavy nuclei~\cite{Ste05b} have allowed us to
constrain the density dependence of the symmetry energy at
sub-saturation densities within approximately
$31.6(\rho/\rho_0)^{0.69}$ and $31.6(\rho/\rho_0)^{1.05}$ as
labelled by $x=0$ and $x=-1$, respectively, in the lower panel of
Fig.\ref{Pressure-eos}~\cite{Che05a,LiBA05c}. While these
constraints are only valid for sub-saturation densities and still
suffer from some uncertainties, compared to the early situation
they represent a significant progress in the field. Further
progress is expected from both the parity violating electron
scattering experiments~\cite{Hor01} at the Jefferson lab that will
help pin down the low density part of the symmetry energy and
heavy-ion reactions with high energy radioactive beams at several
facilities that will help constrain the high density behavior of
the symmetry energy~\cite{LCK08}.

For many astrophysical studies, the EOS is usually expressed in
terms of the pressure as a function of density and isospin
asymmetry. Shown in Fig.~\ref{Pressure-eos} are the pressures for
two extreme cases: symmetric (upper panel) and pure neutron matter
(lower panel). The green area in the density range of
$2-4.6\rho_0$ is the experimental constraint on the pressure
$P_{0}$ of symmetric nuclear matter extracted by Danielewicz,
Lacey and Lynch from analyzing the collective flow data from
relativistic heavy-ion collisions~\cite{dan02}. It is seen that
results from mean-field calculations using the phenomenological
momentum-dependent (MDI) interaction \cite{bali03}, the
Dirac-Brueckner-Hartree-Fock approach with the Bonn B potential
(DBHF) \cite{krastev06}, and the variational calculations by
Akmal, Pandharipande, and Ravenhall (APR)\cite{apr} are all
consistent with this constraint. For pure neutron matter, its
pressure is $P_{\rm PNM}=P_{0}+\rho^2dE_{\rm sym}/d\rho$ and
depends on the density dependence of nuclear symmetry energy.
Since the constraints on the symmetry energy from terrestrial
laboratory experiments are only available for densities less than
about $1.2\rho_0$ as indicated by the green and red squares in the
lower panel, which is in contrast to the constraint on the EOS of
symmetry nuclear matter that is only available at much higher
densities, the most reliable estimate of the EOS of neutron-rich
matter can thus be obtained by extrapolating the underlying model
EOS for symmetric matter and the symmetry energy in their
respective density ranges to all densities. Shown by the shaded
black area in the lower panel is the resulting best estimate of
the pressure of high density pure neutron matter based on the
predictions from the MDI interaction with x=0 and x=-1 as the
lower and upper bounds on the symmetry energy and the
flow-constrained EOS of symmetric nuclear matter. As one expects
and consistent with the estimate in Ref.~\cite{dan02}, the
estimated error bars of the high density pure neutron matter EOS
is much wider than the uncertainty range of the EOS of symmetric
nuclear matter. For the four interactions indicated in the figure,
their predicted EOS's cannot be distinguished by the estimated
constraint on the high density pure neutron matter. In the
following, the astrophysical consequences of this partially
constrained EOS of neutron-rich matter on the mass-radius
correlation, moment of inertia, and the elliptical deformation and
gravitational radiation of (rapidly) rotating neutron stars are
briefly discussed. More details of our studies on these topics can
be found in Refs.~\cite{Kra08b,LiBA06a,Kra07a,Kra07b,Wor07}.

\section{Nuclear constraints on the mass-radius correlation,
moment of inertia, elliptical deformation and gravitational
radiation of rapidly rotating neutron stars}

\begin{figure}[h]
\centering
\includegraphics[totalheight=2.in]{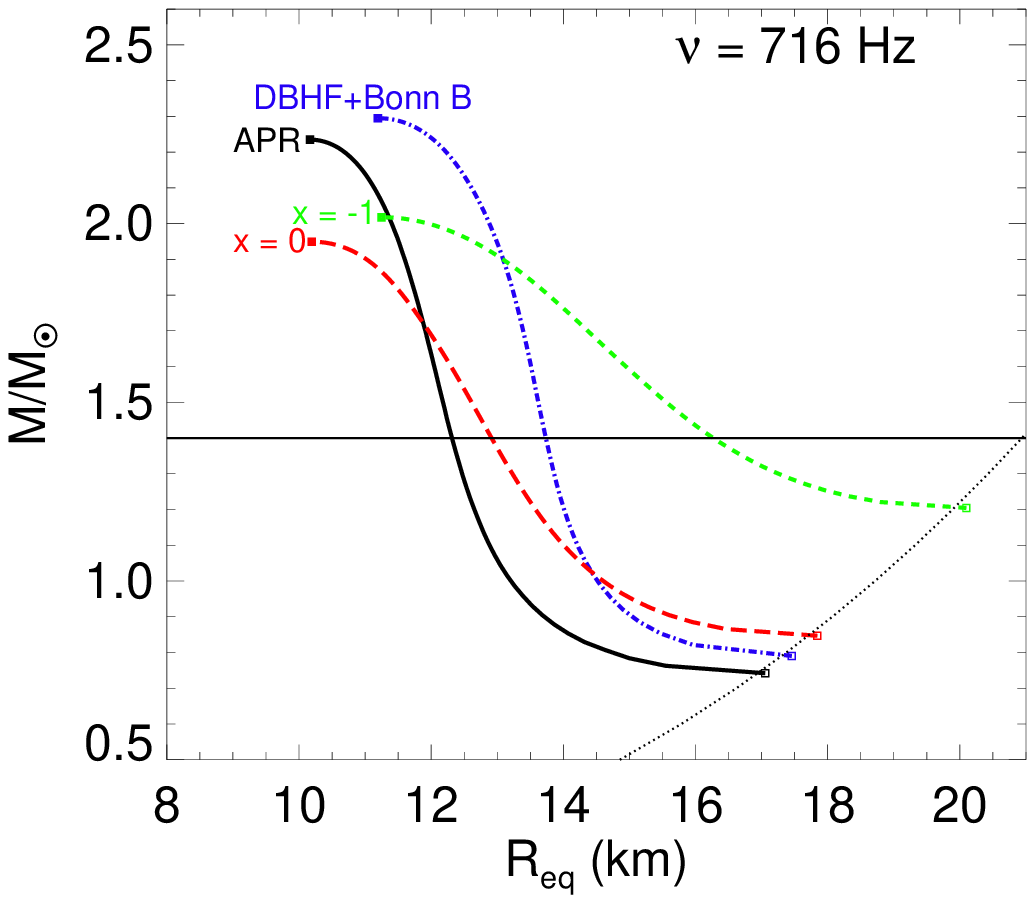}
\includegraphics[totalheight=2.in]{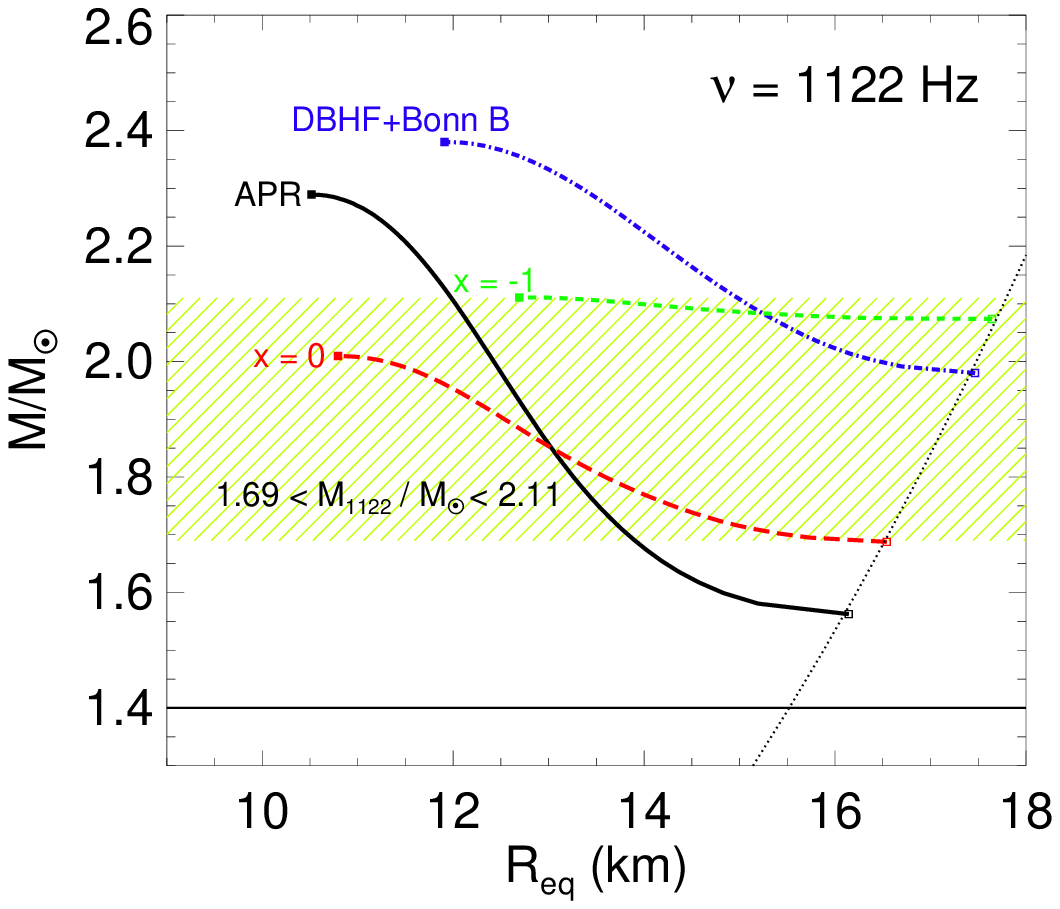}
\caption{(Color online) Gravitational mass versus equatorial radius
for neutron stars rotating at $\nu=716~{\rm Hz}$ and $\nu=1122~{\rm
Hz}$. Taken from Ref.~\cite{Kra07b}.}\label{radii}
\end{figure}

The partially constrained EOS of neutron-rich nuclear matter has
important ramifications on properties of neutron stars. As a first
example, in  Fig.~\ref{radii} we show the mass-radius correlations
for the two fastest rotating neutron stars known as of today.
These pulsars spin at 716~\cite{Hes06} and 1122 Hz~\cite{Kaa06},
respectively. However, based only on the observational data
available so far, their properties have not yet been fully
understood. The analysis of their properties based on the EOS and
symmetry energy constrained by the terrestrial laboratory data is
thus especially interesting. Setting the observed frequency of the
pulsar as the Kepler frequency, corresponding to the highest
possible frequency for a star before its starts to shed mass at
the equator, one can obtain an estimate of its maximum radius as a
function of mass $M$,
\begin{eqnarray}\label{eq.17}
R_{\rm max}(M)=\chi\left(\frac{M}{1.4M_{\odot}}\right)^{1/3}~{\rm
km},
\end{eqnarray}
with $\chi=20.94$ for rotational frequency $\nu=716~{\rm Hz}$ and
$\chi=15.52$ for $\nu=1122~{\rm Hz}$. The maximum radii are shown
with the dotted lines in Fig.~\ref{radii}. It is seen that the
range of allowed masses supported by a given EOS for rapidly
rotating neutron stars becomes narrower than the one of static
configurations. This effect becomes stronger with increasing
frequency and depends upon the EOS. Since predictions from the
$x=0$ and $x=-1$ EOSs represent the limits of the neutron star
models consistent with the experimental data from terrestrial
nuclear laboratories, one can predict that the mass of the neutron
star rotating at 1122 Hz is between 1.7 and $2.1$ solar
mass~\cite{Kra07b}.

\begin{figure}[!t]
\centering
\includegraphics[totalheight=4.7cm]{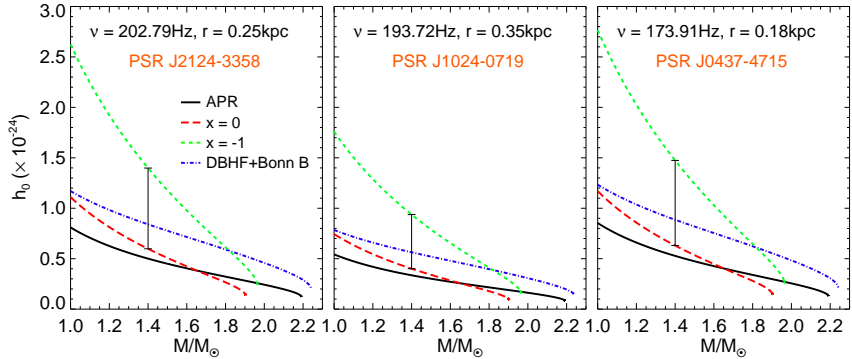}
\vspace{5mm} \caption{Gravitational-wave strain amplitude as a
function of the neutron star mass. The error bars between the $x=0$
and $x=-1$ EOSs provide a limit on the strain amplitude of the
gravitational waves to be expected from these neutron stars, and
show a specific case for stellar models of $1.4M_{\odot}$. Taken
from ref.\cite{Kra08b}.} \label{GWs}
\end{figure}

Another interesting example is the gravitational radiation
expected from elliptically deformed pulsars. Gravitational waves
(GWs) are tiny disturbances in space-time and are a fundamental,
although not yet directly confirmed, prediction of General
Relativity. Gravitational wave astrophysics would open an entirely
new non-electromagnetic window to the Cosmos, making it possible
to probe physics that is hidden or dark to current electromagnetic
observations~\cite{Flanagan:2005yc}. Elliptically deformed pulsars
are among the primary possible sources of the GWs. Very recently
the LIGO and GEO collaborations have set upper limits on the GWs
expected from 78 radio pulasrs~\cite{Ligo}. Gravitational waves
are characterized by a strain amplitude $h_0$ which can be written
as
\begin{equation}\label{Eq.4}
h_0=\chi\frac{\Phi_{22}\nu^2}{r},
\end{equation}
with $\chi=\sqrt{2048\pi^5/15}G/c{^4}$. In the above equation, $r$
is the distance between the pulsar and the detector, and the
$\Phi_{22}$ is the quadrupole moment of the mass distribution. For
slowly rotating neutron stars, one has~\cite{Owen:2005PRL}
\begin{equation}\label{Eq.3}
\Phi_{22,max}=2.4\times
10^{38}g\hspace{1mm}cm^2\left(\frac{\sigma}{10^{-2}}\right)
\left(\frac{R}{10km}\right)^{6.26}
\left(\frac{1.4M_{\odot}}{M}\right)^{1.2}.
\end{equation}
In the above expression, $\sigma$ is the breaking strain of the
neutron star crust which is rather uncertain at present time and
lies in the wide range
$\sigma=[10^{-5}-10^{-2}]$~\cite{HAJS:2007PRL}. In our estimate,
we use the maximum breaking strength, i.e. $\sigma=10^{-2}$. In
Fig.~\ref{GWs} we display the GW strain amplitude, $h_0$, as a
function of stellar mass for three selected millisecond pulsars
which are relatively close to Earth ($r<0.4kpc$) and have
rotational frequencies below $300$ Hz. It is interesting to note
that the predicted $h_0$ is above the design sensitivity of LIGO
detector. The error bars in Fig.~\ref{GWs} between the $x=0$ and
$x=-1$ EOSs provide a constraint on the {\it maximal} strain
amplitude of the gravitational waves emitted by the millisecond
pulsars considered here. The specific case shown in the figure is
for neutron star models of $1.4M_{\odot}$. Depending on the exact
rotational frequency, distance to detector, and details of the
EOS, the {\it maximal} $h_0$ is in the range $\sim[0.4-1.5]\times
10^{-24}$. These estimates do not take into account the
uncertainties in the distance measurements. They also should be
regarded as upper limits since the quadrupole moment
(Eq.~(\ref{Eq.3})) has been calculated with $\sigma=10^{-2}$
(where $\sigma$ can go as low as $10^{-5}$).

To emit GWs a pulsar must have a quadrupole deformation. The latter
is normally characterized by the ellipticity which is related to the
neutron star maximum quadrupole moment $\Phi_{22}$ and the moment of
inertia via~\cite{Owen:2005PRL}
\begin{equation}\label{Eq.2}
\epsilon = \sqrt{\frac{8\pi}{15}}\frac{\Phi_{22}}{I_{zz}},
\end{equation}
For slowly rotating neutron stars, one can use the following
empirical relation \cite{Lattimer:2005}
\begin{equation}\label{Eq.5}
I_{zz}\approx (0.237\pm
0.008)MR^2\left[1+4.2\frac{Mkm}{M_{\odot}R}
+90\left(\frac{Mkm}{M_{\odot}R}\right)^4\right]
\end{equation}
This expression is shown to hold for a wide class of equations of
state which do not exhibit considerable softening and for neutron
star models with masses above $1M_{\odot}$~\cite{Lattimer:2005}.

\begin{figure}[t!]
\centering
\includegraphics[height=4.7cm]{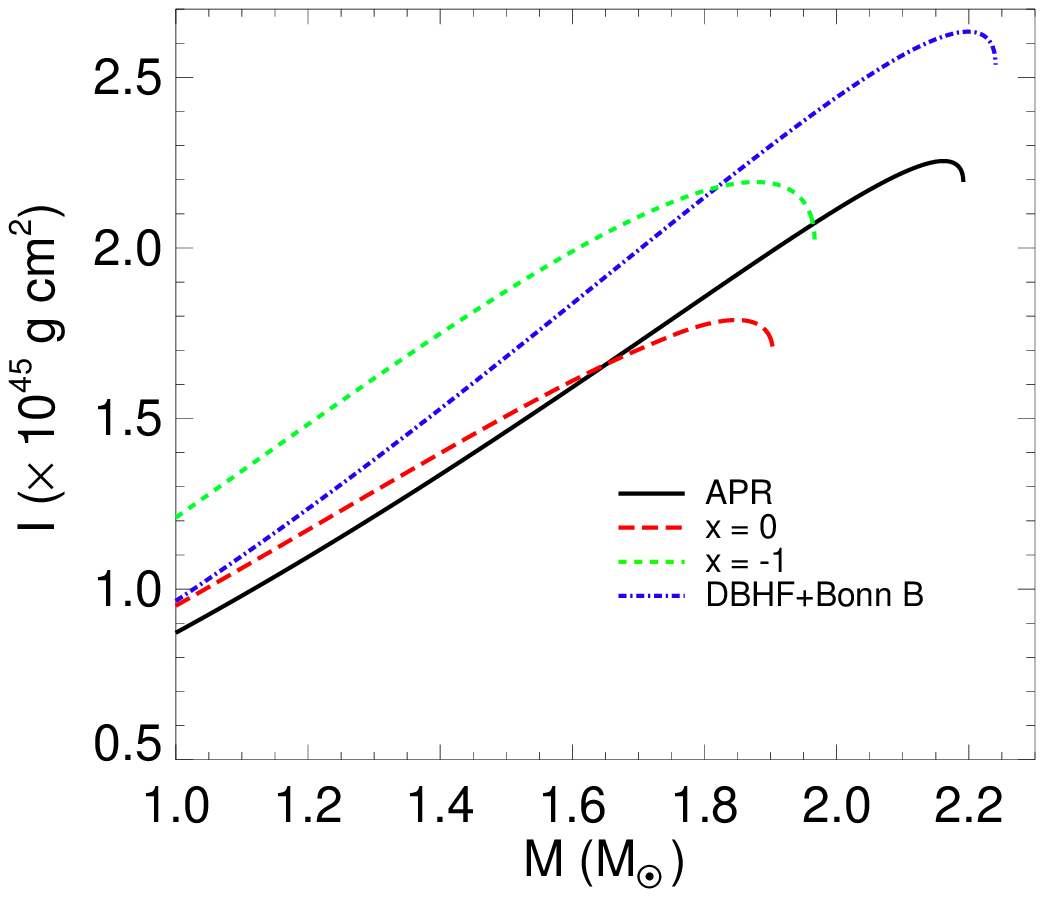}
\includegraphics[height=4.7cm]{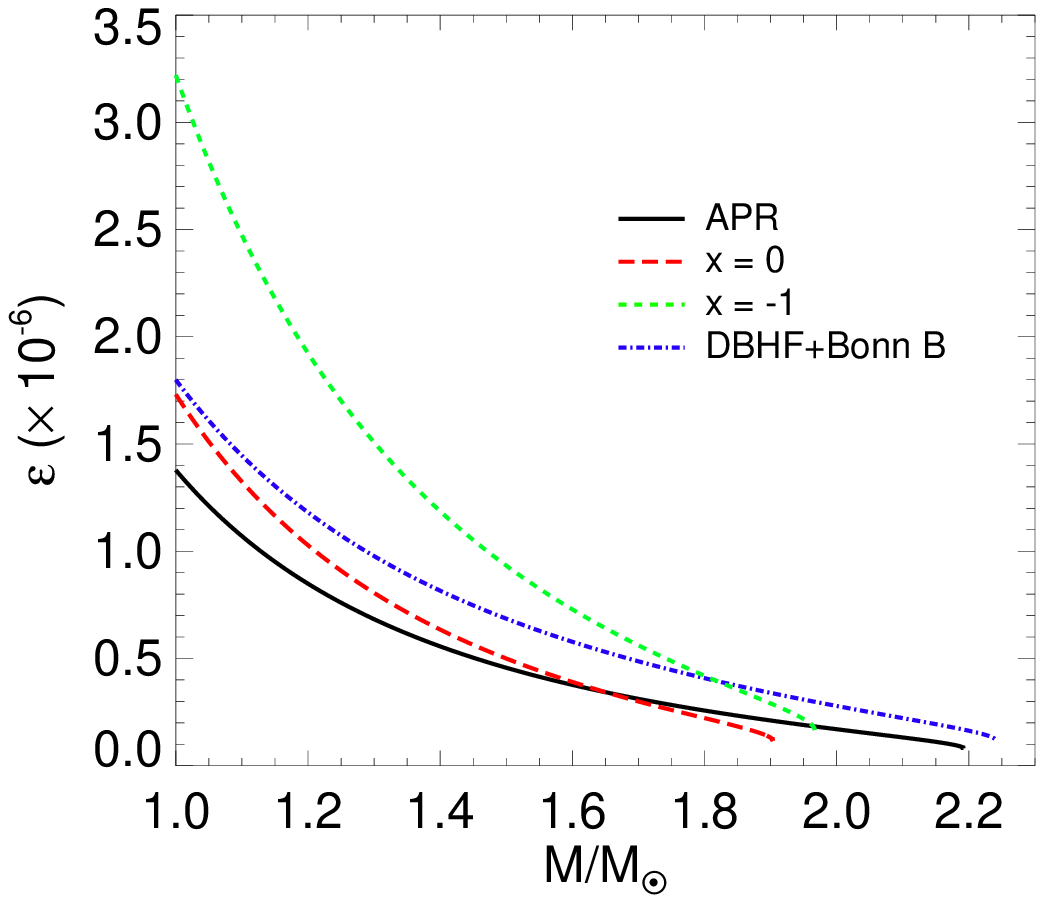}
\caption{Neutron star moment of inertia (left panel) and Ellipticity
(right panel). Taken from ref.\cite{Kra08b}.}\label{iel}
\end{figure}

Fig.~\ref{iel} displays the neutron star moment of inertia (left
panel) and ellipticity (right panel). It is interesting to mention
that a fuducial value of $I_{zz}=10^{45}$g cm$^2$ is normally
assumed in the literature. Our calculations indicate that the
$I_{zz}$ is strongly mass dependent. This observation is
consistent with previous calculations. Moreover, the ellipticity
decreases with increasing mass. The magnitude is above the lowest
upper limit of $4\times 10^{-7}$ estimated for the PSR
J2124-3358~\cite{Ligo}. Interestingly, essentially all obervables
depend strongly on the EOS of neutron-rich matter. In particular,
the MDI EOSs, adopting the same symmetric matter EOS but different
density dependence of the symmetry energy, sets useful nuclear
boundaries for these gravitational wave observables.

In summary, the heavy-ion physics community has made significant
progress in constraining the EOS of neutron-rich nuclear matter in
recent years. In particular, comprehensive analyses of several
isospin effects including the isospin diffusion and isoscaling in
heavy-ion reactions and the size of neutron skins in heavy nuclei
have allowed us to constrain the symmetry energy at sub-saturation
densities within approximately $31.6(\rho/\rho_0)^{0.69}$ and
$31.6(\rho/\rho_0)^{1.05}$. While the currently existing data only
allowed us to constrain the symmetry energy and thus the EOS of
neutron-rich matter in a narrow range, it can already help to put
some useful constraints on several interesting observables in
astrophysics, such as the mass-radius correlation, moment of
inertia, and the elliptical deformation and gravitational
radiation of (rapidly) rotating neutron stars. With the parity
violating electron scattering experiments and heavy-ion reactions
with high energy radioactive beams, it will be possible in the
future to map out accurately the entire density dependence of the
symmetry energy.

\section*{Acknowledgments}
This work was supported in part by the US National Science
Foundation under Grant No. PHY-0652548, PHY-0757839 and PHY-0457265, the
Research Corporation under Award No. 7123, the Advanced Research
Program of the Texas Coordinating Board of Higher Education under
grant no. 003565-0004-2007, the Welch Foundation under Grant No.
A-1358, the National Natural Science Foundation of China under
Grant Nos. 10575071 and 10675082, MOE of China under project
NCET-05-0392, Shanghai Rising-Star Program under Grant
No.06QA14024, the SRF for ROCS, SEM of China, and the National
Basic Research Program of China (973 Program) under Contract No.
2007CB815004.

\vfill\eject
\end{document}